\documentclass[aps,prd,secnumarabic, nobibnotes, twocolumn,superscriptaddress]{revtex4-1}

\usepackage{amsfonts}
\usepackage{mathrsfs}
\usepackage{amsmath}
\usepackage{color}
\usepackage{natbib}
\usepackage{graphicx}
\usepackage{bm}
\usepackage{amssymb}
\usepackage{xspace}
\usepackage{epstopdf}
\usepackage{dcolumn}
\usepackage{longtable}
\usepackage{multirow}
\usepackage[colorlinks=true, letterpaper=true, pdfstartview=FitV, linkcolor=blue, citecolor=blue, urlcolor=blue]{hyperref}

\begin{document}

\title{Trigonal warping induced terraced  spin texture and nearly  perfect spin polarization in graphene with Rashba effect}


\author{Da-Shuai Ma}
\affiliation{Beijing Key Laboratory of Nanophotonics and Ultrafine Optoelectronic
Systems, School of Physics, Beijing Institute of Technology, Beijing 100081, China}

\author{Zhi-Ming Yu}
\email{zhiming\_yu@sutd.edu.sg}
\affiliation{Research Laboratory for Quantum Materials, Singapore University of Technology and Design, Singapore 487372, Singapore}

\author{Hui Pan}
\affiliation{Department of Physics and Key Laboratory of Micro-Nano Measurement-Manipulation and Physics (Ministry of Education), Beihang University, Beijing 100191, China}

\author{Yugui Yao}
\email{ygyao@bit.edu.cn}
\affiliation{Beijing Key Laboratory of Nanophotonics and Ultrafine Optoelectronic
Systems, School of Physics, Beijing Institute of Technology, Beijing 100081, China}

\date{\today}
\begin{abstract}
Electrical tunability of spin polarization has been a focus in spintronics.
Here, we report that the trigonal warping (TW) effect, together with spin-orbit coupling (SOC), can lead to two distinct  magnetoelectric effects in low-dimensional systems.
Taking  graphene with Rashba SOC as example, we study the electronic properties and spin-resolved scattering of system.
It is found that the TW effect  gives rise to a terraced   spin texture in low-energy bands and  can render significant spin polarization in the scattering, both resulting in an efficient  electric  control of spin polarization.
Our work unveils not only SOC  but also the TW effect is important for low-dimensional spintronics.

\end{abstract}

\maketitle

\section{Introduction}
Symmetry, the fundamental physics law in solids, means invariance and guarantees certain degeneracy of band structures \cite{ChiuRMP}. For materials with both time reversal symmetry and inversion symmetry, each band is at least double degeneracy and spin neutral. While for spintronics \cite{wolf2001spintronics,vzutic2004spintronics,Pulizzi2012}, a central issue is to break spin neutral and to  produce an efficient control of spin polarization \cite{Pulizzi2012,dyrdal2009spin}. Usually, spin polarization acquires breaking time reversal symmetry \cite{jungwirth2016antiferromagnetic,eichler2017electron,eichler2017electron,yazyev2008magnetic,wang2005programmable,eichler2017Thermal}, such as
applying zeeman field. The exploration of SOC effect  makes breaking spatial symmetry rather than time reversal symmetry to achieve spin polarization possible  and such effect tremendously extends the scope of spintronics' application \cite{LiuPRB2016,engels1997experimental,ohe2005mesoscopic,yamamoto2005spin,Zhao2016}.

The absence of inversion symmetry in two-dimensional material  tends to distort the Fermi surface of system, making the appearance of warping effect in low-energy bands, such as TW in transition-metal dichalcogenides (TMDs) \cite{kormanyos2013monolayer} and graphene (silicene) with Rashba SOC effect \cite{rakyta2010trigonal,yu2015electric}, and hexagonal warping in the surface state of topological insulator \cite{FuPRL,YuPRB2017}.
In most previous studies \cite{yokoyama2013controllable,tsai2013gated,beenakker2008colloquium,liu2012spin,grujic2014spin}, the TW effect is considered as a perturbation and  as being  irrelevant to the main features of system. Recently, T. Habe et. al. predicted that the spatial  separation of up-spin and down-spin can be easily realized using a atomic step in TMDs \cite{Habe2015}. Surprisedly, it is found the TW effect is the essential element for generating this spin splitter, indicating that the effect of TW in  spintronics has been strongly underestimated in the past. In  Ref. [\onlinecite{Habe2015}],  only the geometric properties of TW effect are used.
While the  direct interaction between TW effect and SOC with real spin is absent, as  spin is a good quantum number there \cite{Habe2015}, guaranteed by the mirror symmetry of system \cite{XiaoPRL2007}. Thus, when TW effect has a direct   influence on real spin, such as TW effect in graphene with Rashba SOC, one can expect more intriguing   magnetoelectric phenomena may emerge.
Particularly,  moderate  Rashba effect has been recently reported in many materials \cite{marchenko2012giant,ishizaka2011giant,eremeev2012ideal,varykhalov2012ir,di2013electric,liebmann2016giant,matetskiy2015two,volobuev2017giant,niesner2016giant}.
Consequently, the investigation of TW effect in Rashba SOC is   not only necessary for a better  understanding of fundamental  physics of Rashba SOC  but also useful for potential application of identified Rashba SOC materials.

In this work, we study  the electronic properties and spin-resolved transport of graphene with moderate  Rashba SOC.
Here, we choose graphene as example because graphene is a typical two-dimensional material with simplest Hamiltonian \cite{KanePRL2005}.
The main physics obtained here can be  applied  to more general cases.
Compared to previous studies without TW effect \cite{Bercioux2010,liu2012spin,grujic2014spin}, we find TW effect induces two overlooked but distinct features, which both can produce efficient electric control of spin polarization.
(i): For the low-energy bands, the Fermi surface is trigonally warped and the spin direction of the electrons residing at the concave segments of Fermi surface are almost same, giving rise to a terraced  spin texture [see Fig. \ref{fig1}(c)]. Such spin texture is distinct from that in other identified SOC materials. Moreover, due to the presence of the plateaus in the terraced  spin texture, one can obtain a current with strong spin polarization by simply applying an electric potential [see Fig. \ref{fig1}(e)].
(ii): When an electron moves through a Rashba barrier [see Fig. \ref{fig2}(a)],  its  spin-resolved transmission probability  would be sensitive to the TW effect.
Without TW effect, the transmitted current is always spin neutral, indicating that Rashba SOC cannot solely generate spin polarization in the scattering \cite{Bercioux2010}. In sharp contrast, we find that when TW effect is taken into account, the transmitted current would be spin polarized, as the transmission probabilities of  up spin and down spin are no longer identical.  Remarkably, by tuning electric potential, one can obtain a nearly perfect spin polarization in the transmitted region.
Moreover, the transmitted electrons with different  spin are collimated to opposite directions, leading to   an electric field controlled spin splitter (see Fig. \ref{fig4}).
Our work unveils that due to TW effect,  Rashba SOC is sufficient to generate current with strong spin polarization, showing a wider scope of potential application of Rashba SOC materials.

\section{Model}

The low-energy electrons of  graphene with Rashba SOC locate  in  the vicinity of  two inequivalent valley points, labeled as $K$ and $K'$.
The two valleys are not independent,  but are connected by time reversal symmetry $\cal{T}$.
Hence, in the following, we will focus on the physical properties of  low-energy electrons residing at  $K$ valley.
The effective  Hamiltonian of system  expanded around  $K$ valley reads \cite{rakyta2010trigonal,yu2015electric}
\begin{eqnarray}
H & = & v_{F}\tau_{0}(k_{x}\sigma_{x}+k_{y}\sigma_{y})-\frac{3}{2}\lambda(\tau_{x}\sigma_{y}+\tau_{y}\sigma_{x})\nonumber \\
 &  & -\frac{\sqrt{3}}{4}a\lambda\left[(k_{x}\tau_{x}+k_{y}\tau_{y})\sigma_{y}-(k_{x}\tau_{y}-k_{y}\tau_{x})\sigma_{x}\right],\label{Cham}
\end{eqnarray}
where $\bm{\tau}$ ($\bm{\sigma}$) is Pauli matrix acting on sublattice (spin) space,  $v_{F}=\sqrt{3}at/2$ is the Fermi velocity with $t$ the hopping parameter and $a$  the lattice constant of graphene,  and $\lambda$ denotes the strength of Rashba SOC.
In the previous works \cite{Bercioux2010,liu2012spin,grujic2014spin}, only the leading term (containing zero order of $k_{x(y)}$) of Rashba SOC effect is keeped and hence the last term in Hamiltonian (\ref{Cham}) generally is omitted.
Such approximate is suitable when Rashba SOC is weak.
However, when the strength of Rashba SOC  becomes comparable with hopping energy, e.g. $\lambda/t>0.1$, the last term in Hamiltonian (\ref{Cham}) can not be discarded and would have  important influences on the  physical properties of system,   as we will discuss later.

\begin{figure}
\includegraphics[width=8.8cm]{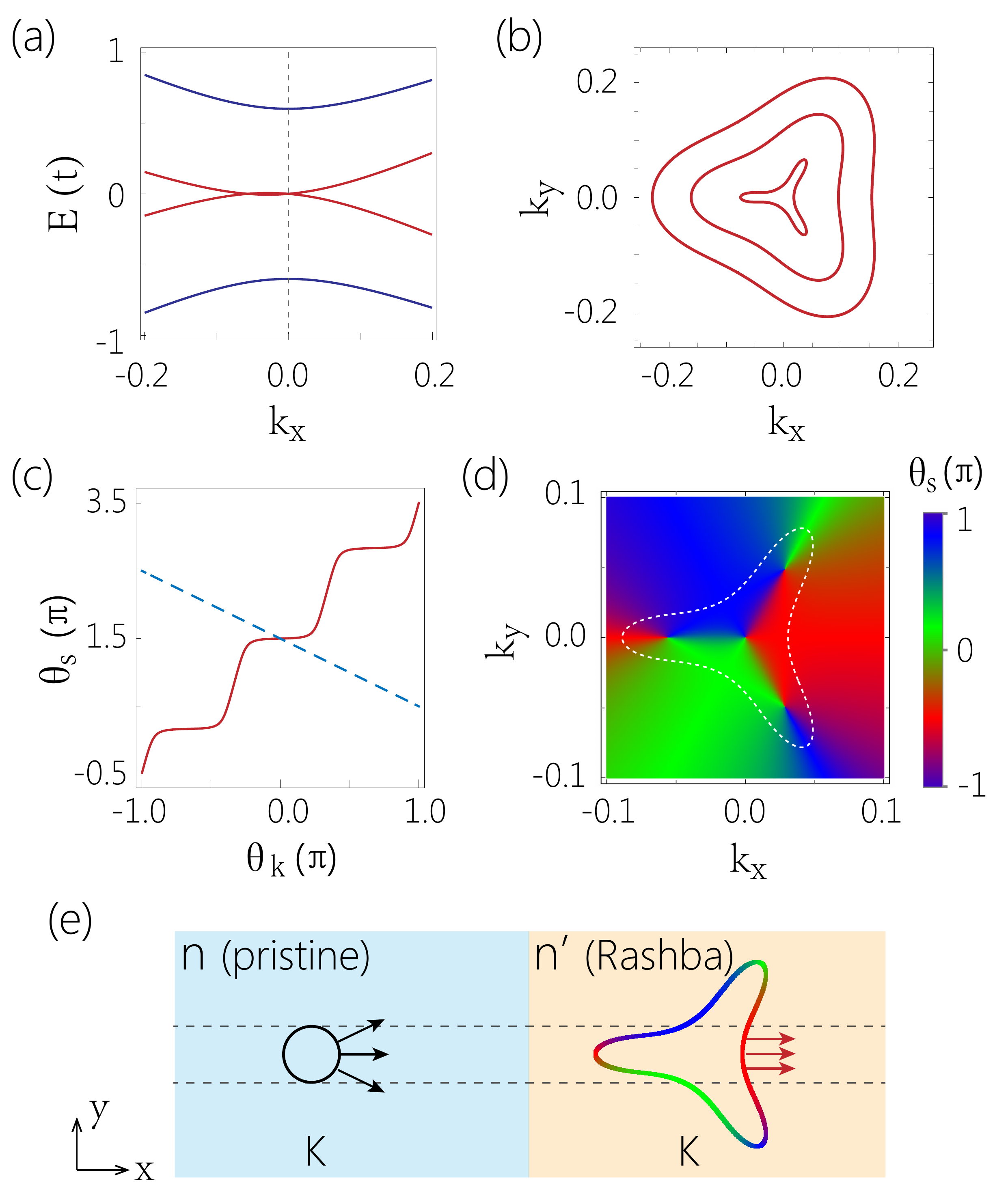}
\caption{(Color online) (a) Band dispersion of graphene with Rashba SOC showing that the two low-energy bands (red lines) are well separated from the other bands (blue lines).
(b) Constant energy contour for the low-energy bands. The TW effect is obvious.
(c) Spin direction $\theta_{s}$  as a function of $\theta_{k}$ for constant energy $E_F=0.02\ t$.
With the TW effect, the spin texture presents a terraced  profile with three separated plateaus  (red solid line). For comparison, we also plot the conventional  spin texture of Rashba system  (blue dashed line).
(d) Spin direction $\theta_{s}$ in momentum space showing the plateaus in (c) persist for  wide momentum   range. The white dashed line is the constant energy contour for $E_{F}=0.02\ t$.
(e) Schematic showing a simple way to generate current with strong spin polarization (only K valley is shown).
Here, we assume $\lambda=0.2\ t$. (a) is calculated with four-band Hamiltonian $H$ (\ref{Cham}), while (b), (c) and (d) are obtained from two-band Hamiltonian $H_w$ (\ref{Wham}).
\label{fig1}}
\end{figure}

Hamiltonian (\ref{Cham}) gives out four bands in momentum space.
With a moderate Rashba SOC ($\lambda=0.2t$),  the two low-energy bands  are well separated from the other two bands in energy space as shown in Fig. \ref{fig1}(a).
Thus, to describe  the low-energy behavior of system,  a two-band model is enough, which can give a clear picture to understand the TW effect.
With a standard process, one can fold Hamiltonian (\ref{Cham}) into a two-band model \cite{yu2015electric}, expressed as
\begin{equation}
H_{w}=-\frac{\sqrt{3}}{2}a\lambda(k_x \sigma_y+k_y \sigma_x)
+\gamma
\left(\begin{array}{cc}
0 & ik_{-}^{2}\\
-ik_{+}^{2} & 0
\end{array}\right),\label{Wham}
\end{equation}
with $\gamma=v_F^2/(3\lambda)$ and $k_{\pm}=k_x\pm ik_y$.
For the two limits $\gamma \ll a\lambda$ ($\lambda\gg t$) and $\gamma \gg a\lambda$ ($\lambda\ll t$), the  Fermi surface of system in both cases are circle while the band dispersion are linear and quadratic, respectively.
When $\lambda$ is comparable with $t$, the competition between the two terms of $H_{w}$  would induce  the trigonally warped Fermi surface.
In Fig. \ref{fig1}(b), we present  the constant energy contour of Hamiltonian $H_{w}$ with $\lambda=0.2t$. One can see the Fermi surface of system  strongly deviates from circle and exhibits obvious TW effect.
Moreover, due to the presence of $\sigma_{x(y)}$, the basis of $H_{w}$ is related to real spin.
Thus, TW term  would have direct  influence on the spin texture of system, which may induce intriguing  phenomena.

Since $H_{w}$ does not contain $\sigma_{z}$, the out-of-plane component of spin $s_{z}\propto\langle\sigma_{z}\rangle$ vanishes.
Then the spin of low-energy electron lies in the plane, and its direction is given as $\theta_{s}=\arg(s_{x}+is_{y})$ with $s_{x(y)}=\langle\sigma_{x(y)}\rangle$.
A straightforward calculation leads to
\begin{equation}
\theta_{s}=\arg(d_{x}+id_{y}),
\end{equation}
with $d_{x(y)}$ the coefficients of $\sigma_{x(y)}$ in Hamiltonian (\ref{Wham}).
When Rashba SOC $\lambda$ dominates the direct hoping energy $t$, one has  $H_{w}\sim k_x \sigma_y+k_y \sigma_x$ and then the spin direction of electron  is normal to its $\bm{k}$ direction as $\theta_{s}=\arg(k_{y}+i k_{x})$, recovering the conventional spin texture of Rashba SOC systems \cite{Manchon}.

\section{Terraced  spin texture}

Interestingly, when $\lambda$ is comparable with $t$ and then the  TW effect becomes obvious [see Fig. \ref{fig1}(b)], the spin texture of system would be very different.
In Fig. \ref{fig1}(c),  we plot the spin direction of electron  ($\theta_{s}$) as a function of  azimuth angle of  momentum $\theta_{k}$ {[}$\theta_{k}\equiv\arg(k_{x}+ik_{y})${]} for constant energy.
Compared to spin texture of conventional Rashba system [corresponding to the case of let $\gamma=0$ in Hamiltonian (\ref{Wham})], one observes that  the TW effect  brings two distinct features into the spin texture.

First, the winding number of   Fermi surface (not very close to zero energy)  is $2$ rather than $1$. Because  when an electron moves around Fermi surface, the variation of spin direction   ($\theta_{s}$) here is $4\pi$ [red solid line in Fig. \ref{fig1}(c)] while that in conventional Rashba system is  $2\pi$ [blue dashed line in Fig. \ref{fig1}(c)].

Second and remarkably,  the spin texture here features a terraced   profile with three separated plateaus as shown in Fig. \ref{fig1}(c).
The three plateaus locate at the three concave segments of Fermi surface, respectively [see Fig. \ref{fig1}(d)]. Thus, for  electrons residing at the concave segments of Fermi surface, they  would  share similar spin direction, indicating  strong spin polarization.
Moreover, the plateaus  and hence the  strong spin polarization persist in a large momentum and energy range as shown in  Fig. \ref{fig1}(d).
This  unique spin texture can not be found in other SOC materials, e.g. the surface state of topological insulator,  Weyl (Dirac) semimetals  and nodal line semimetal \cite{Sheng_JPCL,Souma2011,Sheng2017,CongCaAgBi,Zhang2017}.
Hence, one can expect  it may have distinct  influence on the transport and optical properties of system \cite{ZhouPRB2014}.

A direct application of the terraced spin texture is to generate spin-polarized current. Consider a two-dimensional junction as  shown in Fig. \ref{fig1}(e). The left  ($x<0$)  region is pristine graphene and the right ($x>0$) region is the graphene with Rashba SOC.
The Fermi surface of both sides of junction   can be separately  controlled by the  bias voltage  and electric potential energy.
Hence, with a fine control,  the transmitted electrons in $x>0$ region can be all from one concave segment of Fermi surface [see  Fig. \ref{fig1}(e)], generating  a current with strong spin polarization.
Note that in above discussions,  only $K$ valley is taken into account.
In fact scattering process  simultaneously happens at the $K'$ valley.
However, the transmitted electrons of  $K'$ valley would not reside at a concave segment of Fermi surface  and hence  would weaken the spin polarization of the transmitted current.
Fortunately, valley filter has been experimentally realized in graphene \cite{rycerz2007valley,wu2016full}. Thus by applying  a valley filter, one still can obtain a current with strong spin polarization.

\section{Spin splitter}

\begin{figure}
\includegraphics[width=8cm]{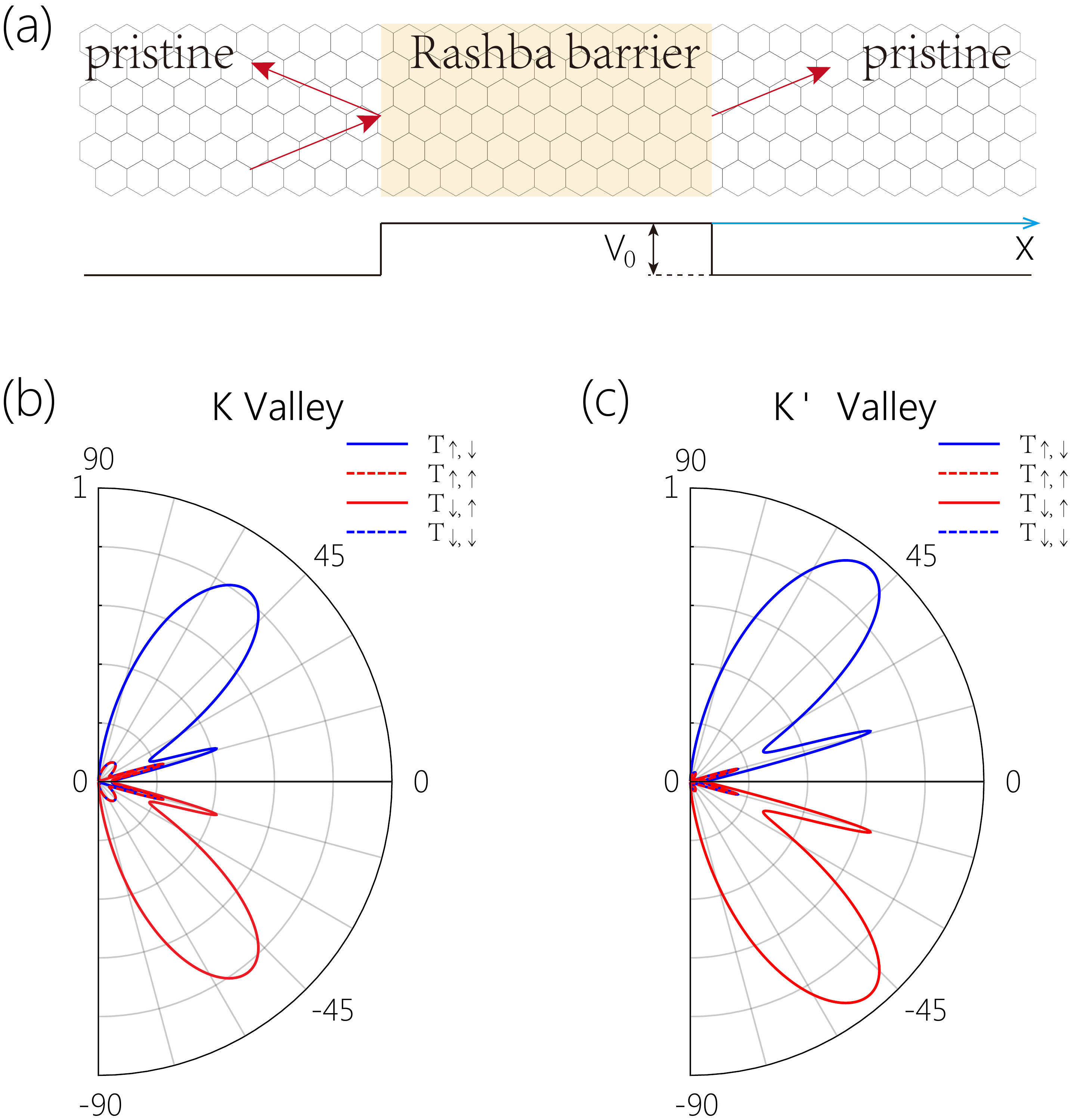}
\caption{(Color online) (a) Schematic figure of a  junction with Rashba barrier.
(b)-(c) Angular plots for transmission probability $T_{s,s^{\prime}}$ as a function of incident angle $\theta_{i}$ for (b)  $K$  and  (c) $K'$ valley.
Here, we assume  $E_{F}=0.12\ t$, $V_{0}=0.1\ t$ and $d=50\ a$.
\label{fig2}}
\end{figure}

Next, we  explore the TW effect of Rashba SOC on the transport properties of system.
Consider a graphene junction as shown in Fig. \ref{fig2}(a), for $x<0$  ($x>d$) the pristine graphene region and for $0<x<d$ the region of graphene with Rashba SOC. To calculate the scattering, we have  to resort to   the four-band Hamiltonian $H$ (\ref{Cham}), as the description of  low-energy physics of pristine graphene requires a four-band model (including spin).
Then, the physics of scattering can be captured by the following model
\begin{eqnarray}
{\cal H}(x) & = & H_{0}[\Theta(-x)+\Theta(x-d)]\nonumber \\
 & + & (H+V_{0})[\Theta(x)-\Theta(x-d)],\label{hami}
\end{eqnarray}
where  $H_0\equiv H(\lambda=0)$ is the Hamiltonian of pristine graphene, $\Theta$ the Heaviside step function and $k_x$ in $H_{(0)}$ is replaced by $-i\partial_{x}$, due to the absence of translation invariance along $x$-direction. Here, we also introduce the electric  potential energy ($V_{0}$) in Rashba region.

In the scattering, the transverse momentum ($k_{y}$) and energy ($E_F$) are conserved while spin can be flipped when electrons pass through Rashba region.
Hence, an incident electron (from $x<0$ region) with certain spin may be reflected or transmitted (into $x>d$ region) as electron with opposite spin.
In addition, since the dispersion of Hamiltonian $H$ (\ref{Cham}) contains quartic terms  of $k_x$, there exist four possible electron states for  given $k_{y}$ and $E_F$.
Then, the typical scattering state of the junction model  reads
\begin{eqnarray}
\Psi(x) & = & \begin{cases}
\psi_{s}^{+}+\sum_{s^{\prime}}r_{s^{\prime}s}\psi_{s^{\prime}}^{-}, & x<0,\\
\sum_{i}c_{i}\psi_{i}^{R}, & 0<x<d\\
\sum_{s^{\prime}}t_{s^{\prime}s}\psi_{s^{\prime}}^{+}, & x>d,
\end{cases},
\end{eqnarray}
where $s(s^{\prime})=\uparrow,\downarrow$ denotes spin, $r_{ss^{\prime}}$ ($t_{ss^{\prime}}$) is the reflection (transmission) amplitude and $c_{i}$ is the scattering amplitude in Rashba region with $i=1,2,3,4$ corresponding to the four scattering  states.
The incident (reflected) wave function is given as $\psi_{\uparrow}^{\pm}=(\bm{0},\varphi_{\pm})^{T}e^{\pm ik_{x}^{+}x+ik_{y}y}$ for $s=\uparrow$ ($s'=\uparrow$) and $\psi_{\downarrow}^{\pm}=(\varphi_{\pm},\bm{0})e^{\pm ik_{x}^{+}x+ik_{y}y}$ for $s=\downarrow$ ($s'=\downarrow$).
Here, $\bm{0}=(0,0)$ is a two-component vector and $\varphi_{\pm}=\frac{1}{\mathcal{N}}(\pm\sqrt{k_{x}^{+2}+k_{y}^{2}},\pm k_{x}^{+}+ik_{y})$
with $k_{x}^{+}=v_{F}^{-1}\sqrt{E-v_{F}^{2}k_{y}^{2}}$ and $\mathcal{N}$  the normalization coefficient. The corresponding
scattering basis states in Rashba region  and in transmitted region  can be  obtained by the conservation of energy and transverse momentum.

The scattering amplitudes can be solved by matching the boundary conditions at two interfaces $x=0$ and $x=d$ (see Appendix \ref{AppA}):
\begin{eqnarray}
\Psi(0^{-})=(1+{\cal U})\Psi(0^{+}), & \ \ \  & \Psi(d^{-})=(1-{\cal U})\Psi(d^{+}),\label{boun}
\end{eqnarray}
with
\begin{equation}
{\cal U}=-i\frac{\sqrt{3}\lambda a}{8v_{F}}(i\tau_{y}\otimes\sigma_{0}+\tau_{x}\otimes\sigma_{z}).
\end{equation}
Though analytical expressions for the scattering amplitudes are difficult to obtain, we can solve them numerically. When the  scattering amplitudes are obtained, one immediately knows the reflection ($R_{s^{\prime}s}$) and transmission ($T_{s^{\prime}s}$) probability, as $R_{s^{\prime}s}=|r_{s^{\prime}s}|^{2}$ and $T_{s^{\prime}s}=|t_{s^{\prime}s}|^{2}$. Due to the presence of mirror symmetry with respect to $y$-direction (${\cal M}_{y}$),
one has \begin{eqnarray}
{\cal M}_{y}{\cal H}(k_{y}){\cal M}_{y}^{-1} & = & {\cal H}(-k_{y}),
\end{eqnarray}
with ${\cal M}_{y}=\tau_{y}\sigma_{x}$ and hence
\begin{eqnarray}
T_{s,s^{\prime}}(k_{y}) & = & T_{-s,-s^{\prime}}(-k_{y}),\label{SS}
\end{eqnarray}
as ${\cal M}_{y}$ not only changes $k_{y}$ to $-k_{y}$ but also flips up (down) spin to down (up) spin.
Meanwhile,  there does not exist additional symmetry to guarantee $T_{s,s^{\prime}}(k_{y})=T_{s,-s^{\prime}}(k_{y})$
or $T_{s,s^{\prime}}(k_{y})=T_{-s,-s^{\prime}}(k_{y})$.
Thus the transmitted current can be spin polarized as shown in Fig. \ref{fig2}(b).
By comparison, in the previous studies of Rashba SOC without TW \cite{Bercioux2010},   equation $T_{s,s^{\prime}}(k_{y})=T_{-s,-s^{\prime}}(k_{y})$
does have been observed due to an artificial emergent symmetry, making transmitted electrons  spin degenerate, in such case, one needs to introduce additional SOC effect (e.g. intrinsic SOC)  to establish spin polarization \cite{Bercioux2010}.
In contrast, we demonstrate here that  when  the TW effect  is included  only Rashba SOC is  enough to achieve spin polarization,
which extremely facilitates the experimental  realization of spin polarization.

\begin{figure}
\includegraphics[width=8.8cm]{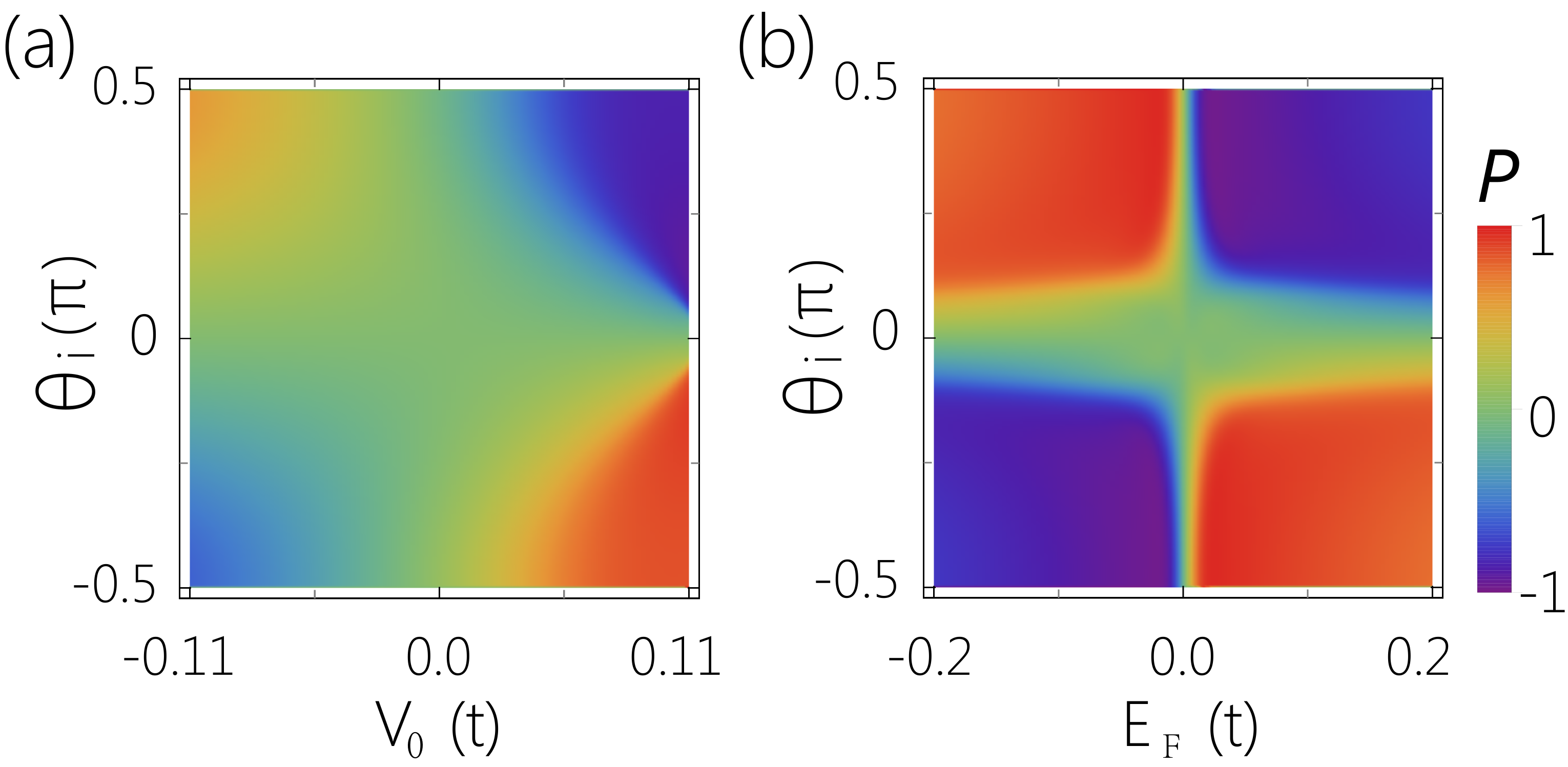}
\caption{(Color online) Total spin polarization $P$ as a function of (a) incident angle $\theta_{i}$ and the height of potential barrier $V_{0}$, and of (b) $\theta_{i}$  and Fermi energy $E_F$.
Here, we set  $E_{F}=0.12\ t$ and $d=50\ a$ in (a), and $V_{0}=E_{F}/1.2$ and $d=50\ a$ in (b).\label{fig3}}
\end{figure}

The above discussion of  symmetry are valid for both valleys. In Fig. \ref{fig2}(b)-(c), we plot the transmission probabilities for both $K$ and $K'$ valley, showing spin polarization of transmitted current can be found in both valleys.
Moreover, we  find that in the scattering, the spin flipping process can dominate the spin preserved  process ($T_{s,-s}\gg T_{s,s}$) and is
 asymmetric between incident angle $\theta_{i}$ {[}e.g. $T_{\uparrow,\downarrow}(\theta_{i})\neq T_{\uparrow,\downarrow}(-\theta_{i})${]}.
Particularly,   $T_{\uparrow,\downarrow}$ would  peak around a certain angle $\theta_{c}^{\nu}$ [see Fig. \ref{fig2}(b) and \ref{fig2}(c)], in the mean time, $T_{\downarrow,\uparrow}$ will peak around $-\theta_{c}^{\nu}$ angle as required by  the mirror symmetry {[}Eq. (\ref{SS}){]}.
Here $\nu=K,K^{\prime}$ denotes the valley.
And, the center angles ($\theta_{c}^{\nu}$) of $T_{\uparrow,\downarrow}$ for the two valleys can be very close (see Fig. \ref{fig2}). Then the transmitted electron with angle around $\theta_{c}^{\nu}$ would exhibit strong spin polarization.
Generally,  one can introduce  \cite{Bercioux2010}
\begin{eqnarray}
P(\theta_{i}) & = & \sum_{\nu=K,K^{\prime}}\frac{T_{\uparrow,\downarrow}^{\nu}+T_{\uparrow,\uparrow}^{\nu}-T_{\downarrow,\downarrow}^{\nu}-T_{\downarrow,\uparrow}^{\nu}}{T_{\uparrow,\downarrow}^{\nu}+T_{\uparrow,\uparrow}^{\nu}+T_{\downarrow,\downarrow}^{\nu}+T_{\downarrow,\uparrow}^{\nu}},
\end{eqnarray}
to quantify the total spin polarization of transmitted electrons.
According to the mirror symmetry ${\cal{M}}_y$,  $P$ should be  antisymmetry with respect to incident angle
\begin{eqnarray}
P(\theta_{i}) & = & -P(-\theta_{i}).\label{ATh}
\end{eqnarray}
Moreover, due to the emergent particle-hole symmetry in Hamiltonian $H$ (\ref{Cham}), one also has
\begin{eqnarray}
P(\theta_i, E_{F}+V_0) & = & P(-\theta_i, -E_{F}-V_0).\label{PHS}
\end{eqnarray}
In Fig. \ref{fig3}, we present the evolution of $P$ varying with potential energy and Fermi energy, which shows the spin polarization
of transmitted current  is strong and features aforementioned  symmetries. Remarkably,  nearly perfect spin polarization (e.g. $|P|>0.9$) can happen  in a wide  $E_{F}$ and $V_{0}$ energy scale.

\begin{figure}[t]
\includegraphics[width=8.5cm]{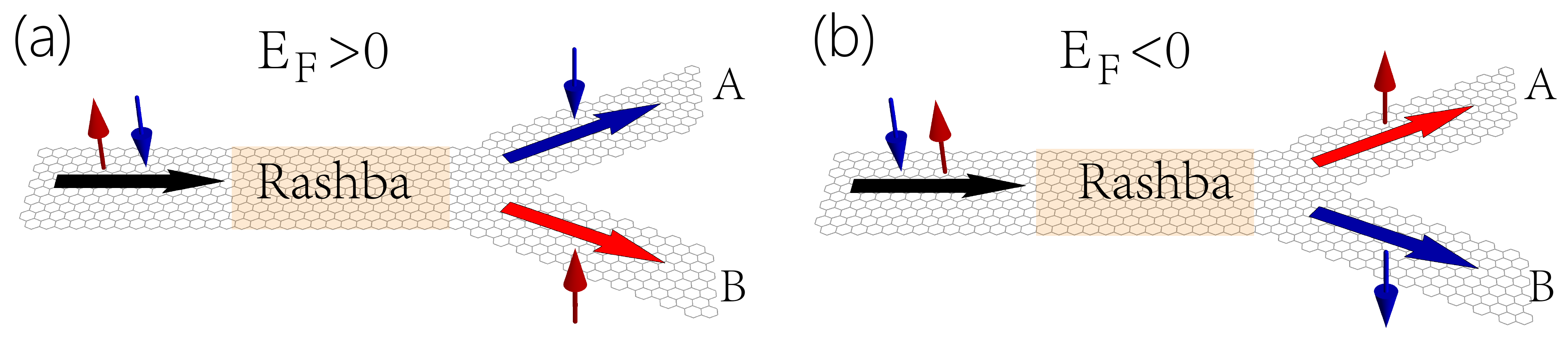}
\caption{(Color online) Schematic figure showing the Y-type structure for  achieving spin splitter. (a) For electron doping ($E_F>0$ and $V_{0}=E_{F}/1.2$), the down-spin (up-spin) of the  current passing through  drain A (B) can dominate up-spin (down-spin). (b) For hole doping ($E_F<0$ and $V_{0}=E_{F}/1.2$), the spin polarization of transmitted current  is switched. \label{fig4}}
\end{figure}

Due to the antisymmetry of $P$ [Eq. (\ref{ATh})], a usual two terminals device is not valid to probe to the polarization effect as an average over incident angle $\theta_{i}$ is involved. Alternatively, one can use a ${\rm Y}$-type structure (see Fig. \ref{fig4}) for splitting transmitted current.
In the ${\rm Y}$-type junction,   the currents moving through drain A and B  can present strong and opposite  spin polarization as guaranteed by Eq. (\ref{ATh}), making the realization of spin splitter.
Also, by changing the sign of $E_F$ and $V_0$ simultaneously, one can switch the sign of $P$  [Eq. (\ref{PHS})] and hence switch the spin polarization of current in drain A and B (see Fig. \ref{fig4}).  In addition, with more delicate setups, e.g. by applying collimators similar to those in electron optics, one could control the incident angle. Then by tuning electrical potential energy or Fermi energy, a nearly perfect spin polarization current could be observed.

\section{Conclusion}

In conclusion, we have predicted two intriguing  features induced by the TW effect in Rashba SOC.
The terraced  spin texture at Fermi surface  predicted here is unique and can not be found in other SOC materials.
Particularly, the spin texture here is related to real spin. Thus, it may be detected by the spin- and angle-resolved photoemission (spinARPES).
In addition, the terraced  spin texture can be used to generate current with strong spin polarization.
We also study the scattering of system with a Rashba barrier and observe strong spin polarization in the transmitted current.
The spin polarization is solely induced by  Rashba SOC and  can be controlled by electric method  (tuning $V_0$ and $E_F$).
And, with a Y-structure junction  (see Fig. \ref{fig4}), one can establish  a spin splitter.
Our work indicates that the warping effect of Fermi surface, together with SOC, can give rise to many interesting phenomena which may be  helpful to the spintronics.

\begin{acknowledgements}
The authors thank P.-T. Xiao for valuable discussions. This work was supported by the MOST Project of China (Grant No. 2014CB920903), the National Key R\&D Program of China (Grant No. 2016YFA0300600), the National Natural Science Foundation of China (Grants No. 11574029, No. 11734003, and No. 11574019), and the Fundamental Research Funds for the Central Universities.
\end{acknowledgements}

\begin{appendix}

\section{Boundary Condition}\label{AppA}

Due to the presence of $\lambda k_{x(y)}$ in Hamiltonian $H$ (\ref{Cham}), the  boundary conditions for the graphene junction with Rashba barrier generally can not be expressed  as the continuum of wavefunction  at interfaces $x=0$ and $x=d$. Here, we establish the boundary conditions of the junction model (\ref{hami}) in detail.

The  junction model (\ref{hami})  in main text with basis $(|A\downarrow\rangle,|B\downarrow\rangle,|A\uparrow\rangle,|B\uparrow\rangle)^{T}$ can be written as
\begin{eqnarray}
{\cal H}(x) & = & \left(\begin{array}{cccc}
0 & v_{F}k_{-} & 0 & h_{0}\\
v_{F}k_{+} & 0 & h_{1} & 0\\
0 & h_{1}^{*} & 0 & v_{F}k_{-}\\
h_{0}^{*} & 0 & v_{F}k_{+} & 0
\end{array}\right),
\end{eqnarray}
where $k_{\pm}=k_{x} \pm ik_{y}$, $\lambda(x)=\lambda \Theta(x)$ and
\begin{eqnarray}
h_{0} & = & 3i\lambda\left(x\right),\\
h_{1} & = & -i\frac{\sqrt{3}}{2}a\lambda\left(x\right)k_{-}.
\end{eqnarray}
The eigenfunction of system is ${\cal H}(x) \Psi\left(x\right)= E  \Psi\left(x\right)$ with $E$ the eigenvalue and $\Psi\left(x\right) =\left(\psi_{1},\ \psi_{2},  \ \psi_{3},\  \psi_{4}  \  \right)^{T}$ the eigenfunction.
Because  the eigenfunction  $\Psi\left(x\right)$ and eigenvalue $E$  are finite, one has
\begin{eqnarray}
\int_{0^{-}}^{0^{+}}{\cal H}(x)\Psi(x)dx & = & \int_{0^{-}}^{0^{+}}E\Psi(x)dx=0,\\
\int_{d+0^{-}}^{d+0^{+}}{\cal H}(x)\Psi(x)dx & = & \int_{d+0^{-}}^{d+0^{+}}E\Psi(x)dx=0.
\end{eqnarray}
The boundary conditions can be established from above equations. A straightforward calculation gives  that   $\psi_{2(3)}(x)$ is  continuum at interfaces ($x=0$ and  $x=d$) while
\begin{eqnarray}
\psi_{1}(0^{+})-\psi_{1}(0^{-}) & = & i\xi \psi_{3}(0),\\
\psi_{4}(0^{+})-\psi_{4}(0^{-}) & = & -i\xi \psi_{2}(0),\\
\psi_{1}(d+0^{+})-\psi_{1}(d+0^{-}) & = & -i\xi \psi_{3}(d),\\
\psi_{4}(d+0^{+})-\psi_{4}(d+0^{-}) & = & i\xi \psi_{2}(d),
\end{eqnarray}
with $\xi=\frac{\sqrt{3}\lambda a}{4v_{F}}$.
The above equations are the boundary conditions of system and the compact forms  have been  given in Eq (\ref{boun}) of main text.
One can check that with such  boundary conditions, current conservation is  satisfied.

\end{appendix}

\bibliography{TW_RSOC_ref}

\begin{thebibliography}{47}%
\makeatletter
\providecommand \@ifxundefined [1]{%
 \@ifx{#1\undefined}
}%
\providecommand \@ifnum [1]{%
 \ifnum #1\expandafter \@firstoftwo
 \else \expandafter \@secondoftwo
 \fi
}%
\providecommand \@ifx [1]{%
 \ifx #1\expandafter \@firstoftwo
 \else \expandafter \@secondoftwo
 \fi
}%
\providecommand \natexlab [1]{#1}%
\providecommand \enquote  [1]{``#1''}%
\providecommand \bibnamefont  [1]{#1}%
\providecommand \bibfnamefont [1]{#1}%
\providecommand \citenamefont [1]{#1}%
\providecommand \href@noop [0]{\@secondoftwo}%
\providecommand \href [0]{\begingroup \@sanitize@url \@href}%
\providecommand \@href[1]{\@@startlink{#1}\@@href}%
\providecommand \@@href[1]{\endgroup#1\@@endlink}%
\providecommand \@sanitize@url [0]{\catcode `\\12\catcode `\$12\catcode
  `\&12\catcode `\#12\catcode `\^12\catcode `\_12\catcode `\%12\relax}%
\providecommand \@@startlink[1]{}%
\providecommand \@@endlink[0]{}%
\providecommand \url  [0]{\begingroup\@sanitize@url \@url }%
\providecommand \@url [1]{\endgroup\@href {#1}{\urlprefix }}%
\providecommand \urlprefix  [0]{URL }%
\providecommand \Eprint [0]{\href }%
\providecommand \doibase [0]{http://dx.doi.org/}%
\providecommand \selectlanguage [0]{\@gobble}%
\providecommand \bibinfo  [0]{\@secondoftwo}%
\providecommand \bibfield  [0]{\@secondoftwo}%
\providecommand \translation [1]{[#1]}%
\providecommand \BibitemOpen [0]{}%
\providecommand \bibitemStop [0]{}%
\providecommand \bibitemNoStop [0]{.\EOS\space}%
\providecommand \EOS [0]{\spacefactor3000\relax}%
\providecommand \BibitemShut  [1]{\csname bibitem#1\endcsname}%
\let\auto@bib@innerbib\@empty
\bibitem [{\citenamefont {Chiu}\ \emph {et~al.}(2016)\citenamefont {Chiu},
  \citenamefont {Teo}, \citenamefont {Schnyder},\ and\ \citenamefont
  {Ryu}}]{ChiuRMP}%
  \BibitemOpen
  \bibfield  {author} {\bibinfo {author} {\bibfnamefont {C.-K.}\ \bibnamefont
  {Chiu}}, \bibinfo {author} {\bibfnamefont {J.~C.~Y.}\ \bibnamefont {Teo}},
  \bibinfo {author} {\bibfnamefont {A.~P.}\ \bibnamefont {Schnyder}}, \ and\
  \bibinfo {author} {\bibfnamefont {S.}~\bibnamefont {Ryu}},\ }\href {\doibase
  10.1103/RevModPhys.88.035005} {\bibfield  {journal} {\bibinfo  {journal}
  {Rev. Mod. Phys.}\ }\textbf {\bibinfo {volume} {88}},\ \bibinfo {pages}
  {035005} (\bibinfo {year} {2016})}\BibitemShut {NoStop}%
\bibitem [{\citenamefont {Wolf}\ \emph {et~al.}(2001)\citenamefont {Wolf},
  \citenamefont {Awschalom}, \citenamefont {Buhrman}, \citenamefont {Daughton},
  \citenamefont {Von~Molnar}, \citenamefont {Roukes}, \citenamefont
  {Chtchelkanova},\ and\ \citenamefont {Treger}}]{wolf2001spintronics}%
  \BibitemOpen
  \bibfield  {author} {\bibinfo {author} {\bibfnamefont {S.}~\bibnamefont
  {Wolf}}, \bibinfo {author} {\bibfnamefont {D.}~\bibnamefont {Awschalom}},
  \bibinfo {author} {\bibfnamefont {R.}~\bibnamefont {Buhrman}}, \bibinfo
  {author} {\bibfnamefont {J.}~\bibnamefont {Daughton}}, \bibinfo {author}
  {\bibfnamefont {S.}~\bibnamefont {Von~Molnar}}, \bibinfo {author}
  {\bibfnamefont {M.}~\bibnamefont {Roukes}}, \bibinfo {author} {\bibfnamefont
  {A.~Y.}\ \bibnamefont {Chtchelkanova}}, \ and\ \bibinfo {author}
  {\bibfnamefont {D.}~\bibnamefont {Treger}},\ }\href
  {http://science.sciencemag.org/content/294/5546/1488} {\bibfield  {journal}
  {\bibinfo  {journal} {Science}\ }\textbf {\bibinfo {volume} {294}},\ \bibinfo
  {pages} {1488} (\bibinfo {year} {2001})}\BibitemShut {NoStop}%
\bibitem [{\citenamefont {{\v{Z}}uti{\'c}}\ \emph {et~al.}(2004)\citenamefont
  {{\v{Z}}uti{\'c}}, \citenamefont {Fabian},\ and\ \citenamefont
  {Sarma}}]{vzutic2004spintronics}%
  \BibitemOpen
  \bibfield  {author} {\bibinfo {author} {\bibfnamefont {I.}~\bibnamefont
  {{\v{Z}}uti{\'c}}}, \bibinfo {author} {\bibfnamefont {J.}~\bibnamefont
  {Fabian}}, \ and\ \bibinfo {author} {\bibfnamefont {S.~D.}\ \bibnamefont
  {Sarma}},\ }\href
  {https://journals.aps.org/rmp/abstract/10.1103/RevModPhys.76.323} {\bibfield
  {journal} {\bibinfo  {journal} {Rev. Mod. Phys.}\ }\textbf {\bibinfo {volume}
  {76}},\ \bibinfo {pages} {323} (\bibinfo {year} {2004})}\BibitemShut
  {NoStop}%
\bibitem [{\citenamefont {Pulizzi}(2012)}]{Pulizzi2012}%
  \BibitemOpen
  \bibfield  {author} {\bibinfo {author} {\bibfnamefont {F.}~\bibnamefont
  {Pulizzi}},\ }\href {\doibase 10.1038/nmat3327} {\bibfield  {journal}
  {\bibinfo  {journal} {Nat. Mater.}\ }\textbf {\bibinfo {volume} {11}},\
  \bibinfo {pages} {367} (\bibinfo {year} {2012})}\BibitemShut {NoStop}%
\bibitem [{\citenamefont {Dyrda{\l}}\ \emph {et~al.}(2009)\citenamefont
  {Dyrda{\l}}, \citenamefont {Dugaev},\ and\ \citenamefont
  {Barna{\'s}}}]{dyrdal2009spin}%
  \BibitemOpen
  \bibfield  {author} {\bibinfo {author} {\bibfnamefont {A.}~\bibnamefont
  {Dyrda{\l}}}, \bibinfo {author} {\bibfnamefont {V.}~\bibnamefont {Dugaev}}, \
  and\ \bibinfo {author} {\bibfnamefont {J.}~\bibnamefont {Barna{\'s}}},\
  }\href {https://journals.aps.org/prb/abstract/10.1103/PhysRevB.80.155444}
  {\bibfield  {journal} {\bibinfo  {journal} {Phys. Rev. B}\ }\textbf {\bibinfo
  {volume} {80}},\ \bibinfo {pages} {155444} (\bibinfo {year}
  {2009})}\BibitemShut {NoStop}%
\bibitem [{\citenamefont {Jungwirth}\ \emph {et~al.}(2016)\citenamefont
  {Jungwirth}, \citenamefont {Marti}, \citenamefont {Wadley}, \citenamefont
  {Wunderlich} \emph {et~al.}}]{jungwirth2016antiferromagnetic}%
  \BibitemOpen
  \bibfield  {author} {\bibinfo {author} {\bibfnamefont {T.}~\bibnamefont
  {Jungwirth}}, \bibinfo {author} {\bibfnamefont {X.}~\bibnamefont {Marti}},
  \bibinfo {author} {\bibfnamefont {P.}~\bibnamefont {Wadley}}, \bibinfo
  {author} {\bibfnamefont {J.}~\bibnamefont {Wunderlich}},  \emph {et~al.},\
  }\href
  {http://www.nature.com/nnano/journal/v11/n3/full/nnano.2016.18.html?foxtrotcallback=true}
  {\bibfield  {journal} {\bibinfo  {journal} {Nat Nanotechnol.}\ }\textbf
  {\bibinfo {volume} {11}},\ \bibinfo {pages} {231} (\bibinfo {year}
  {2016})}\BibitemShut {NoStop}%
\bibitem [{\citenamefont {Eichler}\ \emph {et~al.}(2017)\citenamefont
  {Eichler}, \citenamefont {Sigillito}, \citenamefont {Lyon},\ and\
  \citenamefont {Petta}}]{eichler2017electron}%
  \BibitemOpen
  \bibfield  {author} {\bibinfo {author} {\bibfnamefont {C.}~\bibnamefont
  {Eichler}}, \bibinfo {author} {\bibfnamefont {A.}~\bibnamefont {Sigillito}},
  \bibinfo {author} {\bibfnamefont {S.}~\bibnamefont {Lyon}}, \ and\ \bibinfo
  {author} {\bibfnamefont {J.}~\bibnamefont {Petta}},\ }\href
  {https://journals.aps.org/prl/abstract/10.1103/PhysRevLett.118.037701}
  {\bibfield  {journal} {\bibinfo  {journal} {Phys. Rev. Lett.}\ }\textbf
  {\bibinfo {volume} {118}},\ \bibinfo {pages} {037701} (\bibinfo {year}
  {2017})}\BibitemShut {NoStop}%
\bibitem [{\citenamefont {Yazyev}\ and\ \citenamefont
  {Katsnelson}(2008)}]{yazyev2008magnetic}%
  \BibitemOpen
  \bibfield  {author} {\bibinfo {author} {\bibfnamefont {O.~V.}\ \bibnamefont
  {Yazyev}}\ and\ \bibinfo {author} {\bibfnamefont {M.}~\bibnamefont
  {Katsnelson}},\ }\href
  {https://journals.aps.org/prl/abstract/10.1103/PhysRevLett.100.047209}
  {\bibfield  {journal} {\bibinfo  {journal} {Phys. Rev. Lett.}\ }\textbf
  {\bibinfo {volume} {100}},\ \bibinfo {pages} {047209} (\bibinfo {year}
  {2008})}\BibitemShut {NoStop}%
\bibitem [{\citenamefont {Wang}\ \emph {et~al.}(2005)\citenamefont {Wang},
  \citenamefont {Meng},\ and\ \citenamefont {Wang}}]{wang2005programmable}%
  \BibitemOpen
  \bibfield  {author} {\bibinfo {author} {\bibfnamefont {J.}~\bibnamefont
  {Wang}}, \bibinfo {author} {\bibfnamefont {H.}~\bibnamefont {Meng}}, \ and\
  \bibinfo {author} {\bibfnamefont {J.-P.}\ \bibnamefont {Wang}},\ }\href
  {http://aip.scitation.org/doi/abs/10.1063/1.1857655} {\bibfield  {journal}
  {\bibinfo  {journal} {J. Appl. Polym. Sci.}\ }\textbf {\bibinfo {volume}
  {97}},\ \bibinfo {pages} {10D509} (\bibinfo {year} {2005})}\BibitemShut
  {NoStop}%
\bibitem [{\citenamefont {Pientka}\ \emph {et~al.}(2017)\citenamefont
  {Pientka}, \citenamefont {Waissman}, \citenamefont {Kim},\ and\ \citenamefont
  {Halperin}}]{eichler2017Thermal}%
  \BibitemOpen
  \bibfield  {author} {\bibinfo {author} {\bibfnamefont {F.}~\bibnamefont
  {Pientka}}, \bibinfo {author} {\bibfnamefont {J.}~\bibnamefont {Waissman}},
  \bibinfo {author} {\bibfnamefont {P.}~\bibnamefont {Kim}}, \ and\ \bibinfo
  {author} {\bibfnamefont {B.~I.}\ \bibnamefont {Halperin}},\ }\href
  {https://journals.aps.org/prl/abstract/10.1103/PhysRevLett.119.027601}
  {\bibfield  {journal} {\bibinfo  {journal} {Phys. Rev. Lett.}\ }\textbf
  {\bibinfo {volume} {119}},\ \bibinfo {pages} {027601} (\bibinfo {year}
  {2017})}\BibitemShut {NoStop}%
\bibitem [{\citenamefont {Liu}\ \emph {et~al.}(2016)\citenamefont {Liu},
  \citenamefont {Yu},\ and\ \citenamefont {Liu}}]{LiuPRB2016}%
  \BibitemOpen
  \bibfield  {author} {\bibinfo {author} {\bibfnamefont {D.-P.}\ \bibnamefont
  {Liu}}, \bibinfo {author} {\bibfnamefont {Z.-M.}\ \bibnamefont {Yu}}, \ and\
  \bibinfo {author} {\bibfnamefont {Y.-L.}\ \bibnamefont {Liu}},\ }\href
  {\doibase 10.1103/PhysRevB.94.155112} {\bibfield  {journal} {\bibinfo
  {journal} {Phys. Rev. B}\ }\textbf {\bibinfo {volume} {94}},\ \bibinfo
  {pages} {155112} (\bibinfo {year} {2016})}\BibitemShut {NoStop}%
\bibitem [{\citenamefont {Engels}\ \emph {et~al.}(1997)\citenamefont {Engels},
  \citenamefont {Lange}, \citenamefont {Sch{\"a}pers},\ and\ \citenamefont
  {L{\"u}th}}]{engels1997experimental}%
  \BibitemOpen
  \bibfield  {author} {\bibinfo {author} {\bibfnamefont {G.}~\bibnamefont
  {Engels}}, \bibinfo {author} {\bibfnamefont {J.}~\bibnamefont {Lange}},
  \bibinfo {author} {\bibfnamefont {T.}~\bibnamefont {Sch{\"a}pers}}, \ and\
  \bibinfo {author} {\bibfnamefont {H.}~\bibnamefont {L{\"u}th}},\ }\href
  {https://journals.aps.org/prb/abstract/10.1103/PhysRevB.55.R1958} {\bibfield
  {journal} {\bibinfo  {journal} {Phys. Rev. B}\ }\textbf {\bibinfo {volume}
  {55}},\ \bibinfo {pages} {R1958} (\bibinfo {year} {1997})}\BibitemShut
  {NoStop}%
\bibitem [{\citenamefont {Ohe}\ \emph {et~al.}(2005)\citenamefont {Ohe},
  \citenamefont {Yamamoto}, \citenamefont {Ohtsuki},\ and\ \citenamefont
  {Nitta}}]{ohe2005mesoscopic}%
  \BibitemOpen
  \bibfield  {author} {\bibinfo {author} {\bibfnamefont {J.-i.}\ \bibnamefont
  {Ohe}}, \bibinfo {author} {\bibfnamefont {M.}~\bibnamefont {Yamamoto}},
  \bibinfo {author} {\bibfnamefont {T.}~\bibnamefont {Ohtsuki}}, \ and\
  \bibinfo {author} {\bibfnamefont {J.}~\bibnamefont {Nitta}},\ }\href
  {https://journals.aps.org/prb/abstract/10.1103/PhysRevB.72.041308} {\bibfield
   {journal} {\bibinfo  {journal} {Phys. Rev. B}\ }\textbf {\bibinfo {volume}
  {72}},\ \bibinfo {pages} {041308} (\bibinfo {year} {2005})}\BibitemShut
  {NoStop}%
\bibitem [{\citenamefont {Yamamoto}\ \emph {et~al.}(2005)\citenamefont
  {Yamamoto}, \citenamefont {Ohtsuki},\ and\ \citenamefont
  {Kramer}}]{yamamoto2005spin}%
  \BibitemOpen
  \bibfield  {author} {\bibinfo {author} {\bibfnamefont {M.}~\bibnamefont
  {Yamamoto}}, \bibinfo {author} {\bibfnamefont {T.}~\bibnamefont {Ohtsuki}}, \
  and\ \bibinfo {author} {\bibfnamefont {B.}~\bibnamefont {Kramer}},\ }\href
  {https://journals.aps.org/prb/abstract/10.1103/PhysRevB.72.115321} {\bibfield
   {journal} {\bibinfo  {journal} {Phys. Rev. B}\ }\textbf {\bibinfo {volume}
  {72}},\ \bibinfo {pages} {115321} (\bibinfo {year} {2005})}\BibitemShut
  {NoStop}%
\bibitem [{\citenamefont {Zhao}\ \emph {et~al.}(2016)\citenamefont {Zhao},
  \citenamefont {Liu}, \citenamefont {Yu}, \citenamefont {Quhe}, \citenamefont
  {Zhou}, \citenamefont {Wang}, \citenamefont {Liu}, \citenamefont {Zhong},
  \citenamefont {Han}, \citenamefont {Lu}, \citenamefont {Yao},\ and\
  \citenamefont {Wu}}]{Zhao2016}%
  \BibitemOpen
  \bibfield  {author} {\bibinfo {author} {\bibfnamefont {J.}~\bibnamefont
  {Zhao}}, \bibinfo {author} {\bibfnamefont {H.}~\bibnamefont {Liu}}, \bibinfo
  {author} {\bibfnamefont {Z.}~\bibnamefont {Yu}}, \bibinfo {author}
  {\bibfnamefont {R.}~\bibnamefont {Quhe}}, \bibinfo {author} {\bibfnamefont
  {S.}~\bibnamefont {Zhou}}, \bibinfo {author} {\bibfnamefont {Y.}~\bibnamefont
  {Wang}}, \bibinfo {author} {\bibfnamefont {C.~C.}\ \bibnamefont {Liu}},
  \bibinfo {author} {\bibfnamefont {H.}~\bibnamefont {Zhong}}, \bibinfo
  {author} {\bibfnamefont {N.}~\bibnamefont {Han}}, \bibinfo {author}
  {\bibfnamefont {J.}~\bibnamefont {Lu}}, \bibinfo {author} {\bibfnamefont
  {Y.}~\bibnamefont {Yao}}, \ and\ \bibinfo {author} {\bibfnamefont
  {K.}~\bibnamefont {Wu}},\ }\href {\doibase 10.1016/j.pmatsci.2016.04.001}
  {\bibfield  {journal} {\bibinfo  {journal} {Prog. Mater. Sci.}\ }\textbf
  {\bibinfo {volume} {83}},\ \bibinfo {pages} {24} (\bibinfo {year}
  {2016})}\BibitemShut {NoStop}%
\bibitem [{\citenamefont {Korm{\'a}nyos}\ \emph {et~al.}(2013)\citenamefont
  {Korm{\'a}nyos}, \citenamefont {Z{\'o}lyomi}, \citenamefont {Drummond},
  \citenamefont {Rakyta}, \citenamefont {Burkard},\ and\ \citenamefont
  {Fal'ko}}]{kormanyos2013monolayer}%
  \BibitemOpen
  \bibfield  {author} {\bibinfo {author} {\bibfnamefont {A.}~\bibnamefont
  {Korm{\'a}nyos}}, \bibinfo {author} {\bibfnamefont {V.}~\bibnamefont
  {Z{\'o}lyomi}}, \bibinfo {author} {\bibfnamefont {N.~D.}\ \bibnamefont
  {Drummond}}, \bibinfo {author} {\bibfnamefont {P.}~\bibnamefont {Rakyta}},
  \bibinfo {author} {\bibfnamefont {G.}~\bibnamefont {Burkard}}, \ and\
  \bibinfo {author} {\bibfnamefont {V.~I.}\ \bibnamefont {Fal'ko}},\ }\href
  {https://journals.aps.org/prb/abstract/10.1103/PhysRevB.88.045416} {\bibfield
   {journal} {\bibinfo  {journal} {Phys. Rev. B}\ }\textbf {\bibinfo {volume}
  {88}},\ \bibinfo {pages} {045416} (\bibinfo {year} {2013})}\BibitemShut
  {NoStop}%
\bibitem [{\citenamefont {Rakyta}\ \emph {et~al.}(2010)\citenamefont {Rakyta},
  \citenamefont {Kormanyos},\ and\ \citenamefont
  {Cserti}}]{rakyta2010trigonal}%
  \BibitemOpen
  \bibfield  {author} {\bibinfo {author} {\bibfnamefont {P.}~\bibnamefont
  {Rakyta}}, \bibinfo {author} {\bibfnamefont {A.}~\bibnamefont {Kormanyos}}, \
  and\ \bibinfo {author} {\bibfnamefont {J.}~\bibnamefont {Cserti}},\ }\href
  {https://journals.aps.org/prb/abstract/10.1103/PhysRevB.82.113405} {\bibfield
   {journal} {\bibinfo  {journal} {Phys. Rev. B}\ }\textbf {\bibinfo {volume}
  {82}},\ \bibinfo {pages} {113405} (\bibinfo {year} {2010})}\BibitemShut
  {NoStop}%
\bibitem [{\citenamefont {Yu}\ \emph {et~al.}(2015)\citenamefont {Yu},
  \citenamefont {Pan},\ and\ \citenamefont {Yao}}]{yu2015electric}%
  \BibitemOpen
  \bibfield  {author} {\bibinfo {author} {\bibfnamefont {Z.}~\bibnamefont
  {Yu}}, \bibinfo {author} {\bibfnamefont {H.}~\bibnamefont {Pan}}, \ and\
  \bibinfo {author} {\bibfnamefont {Y.}~\bibnamefont {Yao}},\ }\href
  {https://journals.aps.org/prb/abstract/10.1103/PhysRevB.92.155419} {\bibfield
   {journal} {\bibinfo  {journal} {Phys. Rev. B}\ }\textbf {\bibinfo {volume}
  {92}},\ \bibinfo {pages} {155419} (\bibinfo {year} {2015})}\BibitemShut
  {NoStop}%
\bibitem [{\citenamefont {Fu}(2009)}]{FuPRL}%
  \BibitemOpen
  \bibfield  {author} {\bibinfo {author} {\bibfnamefont {L.}~\bibnamefont
  {Fu}},\ }\href {\doibase 10.1103/PhysRevLett.103.266801} {\bibfield
  {journal} {\bibinfo  {journal} {Phys. Rev. Lett.}\ }\textbf {\bibinfo
  {volume} {103}},\ \bibinfo {pages} {266801} (\bibinfo {year}
  {2009})}\BibitemShut {NoStop}%
\bibitem [{\citenamefont {Yu}\ \emph {et~al.}(2017)\citenamefont {Yu},
  \citenamefont {Ma}, \citenamefont {Pan},\ and\ \citenamefont
  {Yao}}]{YuPRB2017}%
  \BibitemOpen
  \bibfield  {author} {\bibinfo {author} {\bibfnamefont {Z.-M.}\ \bibnamefont
  {Yu}}, \bibinfo {author} {\bibfnamefont {D.-S.}\ \bibnamefont {Ma}}, \bibinfo
  {author} {\bibfnamefont {H.}~\bibnamefont {Pan}}, \ and\ \bibinfo {author}
  {\bibfnamefont {Y.}~\bibnamefont {Yao}},\ }\href {\doibase
  10.1103/PhysRevB.96.125152} {\bibfield  {journal} {\bibinfo  {journal} {Phys.
  Rev. B}\ }\textbf {\bibinfo {volume} {96}},\ \bibinfo {pages} {125152}
  (\bibinfo {year} {2017})}\BibitemShut {NoStop}%
\bibitem [{\citenamefont {Yokoyama}(2013)}]{yokoyama2013controllable}%
  \BibitemOpen
  \bibfield  {author} {\bibinfo {author} {\bibfnamefont {T.}~\bibnamefont
  {Yokoyama}},\ }\href
  {https://journals.aps.org/prb/abstract/10.1103/PhysRevB.87.241409} {\bibfield
   {journal} {\bibinfo  {journal} {Phys. Rev. B}\ }\textbf {\bibinfo {volume}
  {87}},\ \bibinfo {pages} {241409} (\bibinfo {year} {2013})}\BibitemShut
  {NoStop}%
\bibitem [{\citenamefont {Tsai}\ \emph {et~al.}(2013)\citenamefont {Tsai},
  \citenamefont {Huang}, \citenamefont {Chang}, \citenamefont {Lin},
  \citenamefont {Jeng},\ and\ \citenamefont {Bansil}}]{tsai2013gated}%
  \BibitemOpen
  \bibfield  {author} {\bibinfo {author} {\bibfnamefont {W.-F.}\ \bibnamefont
  {Tsai}}, \bibinfo {author} {\bibfnamefont {C.-Y.}\ \bibnamefont {Huang}},
  \bibinfo {author} {\bibfnamefont {T.-R.}\ \bibnamefont {Chang}}, \bibinfo
  {author} {\bibfnamefont {H.}~\bibnamefont {Lin}}, \bibinfo {author}
  {\bibfnamefont {H.-T.}\ \bibnamefont {Jeng}}, \ and\ \bibinfo {author}
  {\bibfnamefont {A.}~\bibnamefont {Bansil}},\ }\href
  {https://www.nature.com/articles/ncomms2525} {\bibfield  {journal} {\bibinfo
  {journal} {Nat. Commun.}\ }\textbf {\bibinfo {volume} {4}},\ \bibinfo {pages}
  {1500} (\bibinfo {year} {2013})}\BibitemShut {NoStop}%
\bibitem [{\citenamefont {Beenakker}(2008)}]{beenakker2008colloquium}%
  \BibitemOpen
  \bibfield  {author} {\bibinfo {author} {\bibfnamefont {C.}~\bibnamefont
  {Beenakker}},\ }\href
  {https://journals.aps.org/rmp/abstract/10.1103/RevModPhys.80.1337} {\bibfield
   {journal} {\bibinfo  {journal} {Rev. Mod. Phys.}\ }\textbf {\bibinfo
  {volume} {80}},\ \bibinfo {pages} {1337} (\bibinfo {year}
  {2008})}\BibitemShut {NoStop}%
\bibitem [{\citenamefont {Tombros}\ \emph {et~al.}(2012)\citenamefont
  {Tombros}, \citenamefont {Jozsa}, \citenamefont {Popinciuc}, \citenamefont
  {Jonkman},\ and\ \citenamefont {van Wees}}]{liu2012spin}%
  \BibitemOpen
  \bibfield  {author} {\bibinfo {author} {\bibfnamefont {N.}~\bibnamefont
  {Tombros}}, \bibinfo {author} {\bibfnamefont {C.}~\bibnamefont {Jozsa}},
  \bibinfo {author} {\bibfnamefont {M.}~\bibnamefont {Popinciuc}}, \bibinfo
  {author} {\bibfnamefont {H.~T.}\ \bibnamefont {Jonkman}}, \ and\ \bibinfo
  {author} {\bibfnamefont {B.~J.}\ \bibnamefont {van Wees}},\ }\href
  {https://journals.aps.org/prb/abstract/10.1103/PhysRevB.85.085406} {\bibfield
   {journal} {\bibinfo  {journal} {Phys. Rev. B}\ }\textbf {\bibinfo {volume}
  {85}},\ \bibinfo {pages} {085406} (\bibinfo {year} {2012})}\BibitemShut
  {NoStop}%
\bibitem [{\citenamefont {Gruji{\'c}}\ \emph {et~al.}(2014)\citenamefont
  {Gruji{\'c}}, \citenamefont {Tadi{\'c}},\ and\ \citenamefont
  {Peeters}}]{grujic2014spin}%
  \BibitemOpen
  \bibfield  {author} {\bibinfo {author} {\bibfnamefont {M.~M.}\ \bibnamefont
  {Gruji{\'c}}}, \bibinfo {author} {\bibfnamefont {M.~{\v{Z}}.}\ \bibnamefont
  {Tadi{\'c}}}, \ and\ \bibinfo {author} {\bibfnamefont {F.~M.}\ \bibnamefont
  {Peeters}},\ }\href
  {https://journals.aps.org/prl/abstract/10.1103/PhysRevLett.113.046601}
  {\bibfield  {journal} {\bibinfo  {journal} {Phys. Rev. Lett.}\ }\textbf
  {\bibinfo {volume} {113}},\ \bibinfo {pages} {046601} (\bibinfo {year}
  {2014})}\BibitemShut {NoStop}%
\bibitem [{\citenamefont {Habe}\ and\ \citenamefont
  {Koshino}(2015)}]{Habe2015}%
  \BibitemOpen
  \bibfield  {author} {\bibinfo {author} {\bibfnamefont {T.}~\bibnamefont
  {Habe}}\ and\ \bibinfo {author} {\bibfnamefont {M.}~\bibnamefont {Koshino}},\
  }\href {\doibase 10.1103/PhysRevB.91.201407} {\bibfield  {journal} {\bibinfo
  {journal} {Phys. Rev. B}\ }\textbf {\bibinfo {volume} {91}},\ \bibinfo
  {pages} {201407} (\bibinfo {year} {2015})}\BibitemShut {NoStop}%
\bibitem [{\citenamefont {Xiao}\ \emph {et~al.}(2007)\citenamefont {Xiao},
  \citenamefont {Yao},\ and\ \citenamefont {Niu}}]{XiaoPRL2007}%
  \BibitemOpen
  \bibfield  {author} {\bibinfo {author} {\bibfnamefont {D.}~\bibnamefont
  {Xiao}}, \bibinfo {author} {\bibfnamefont {W.}~\bibnamefont {Yao}}, \ and\
  \bibinfo {author} {\bibfnamefont {Q.}~\bibnamefont {Niu}},\ }\href {\doibase
  10.1103/PhysRevLett.99.236809} {\bibfield  {journal} {\bibinfo  {journal}
  {Phys. Rev. Lett.}\ }\textbf {\bibinfo {volume} {99}},\ \bibinfo {pages}
  {236809} (\bibinfo {year} {2007})}\BibitemShut {NoStop}%
\bibitem [{\citenamefont {Marchenko}\ \emph {et~al.}(2012)\citenamefont
  {Marchenko}, \citenamefont {Varykhalov}, \citenamefont {Scholz},
  \citenamefont {Bihlmayer}, \citenamefont {Rashba}, \citenamefont {Rybkin},
  \citenamefont {Shikin},\ and\ \citenamefont {Rader}}]{marchenko2012giant}%
  \BibitemOpen
  \bibfield  {author} {\bibinfo {author} {\bibfnamefont {D.}~\bibnamefont
  {Marchenko}}, \bibinfo {author} {\bibfnamefont {A.}~\bibnamefont
  {Varykhalov}}, \bibinfo {author} {\bibfnamefont {M.}~\bibnamefont {Scholz}},
  \bibinfo {author} {\bibfnamefont {G.}~\bibnamefont {Bihlmayer}}, \bibinfo
  {author} {\bibfnamefont {E.}~\bibnamefont {Rashba}}, \bibinfo {author}
  {\bibfnamefont {A.}~\bibnamefont {Rybkin}}, \bibinfo {author} {\bibfnamefont
  {A.}~\bibnamefont {Shikin}}, \ and\ \bibinfo {author} {\bibfnamefont
  {O.}~\bibnamefont {Rader}},\ }\href
  {https://www.nature.com/articles/ncomms2227} {\bibfield  {journal} {\bibinfo
  {journal} {Nat. Commun.}\ }\textbf {\bibinfo {volume} {3}},\ \bibinfo {pages}
  {1232} (\bibinfo {year} {2012})}\BibitemShut {NoStop}%
\bibitem [{\citenamefont {Ishizaka}\ \emph {et~al.}(2011)\citenamefont
  {Ishizaka}, \citenamefont {Bahramy}, \citenamefont {Murakawa}, \citenamefont
  {Sakano}, \citenamefont {Shimojima}, \citenamefont {Sonobe}, \citenamefont
  {Koizumi}, \citenamefont {Shin}, \citenamefont {Miyahara}, \citenamefont
  {Kimura} \emph {et~al.}}]{ishizaka2011giant}%
  \BibitemOpen
  \bibfield  {author} {\bibinfo {author} {\bibfnamefont {K.}~\bibnamefont
  {Ishizaka}}, \bibinfo {author} {\bibfnamefont {M.}~\bibnamefont {Bahramy}},
  \bibinfo {author} {\bibfnamefont {H.}~\bibnamefont {Murakawa}}, \bibinfo
  {author} {\bibfnamefont {M.}~\bibnamefont {Sakano}}, \bibinfo {author}
  {\bibfnamefont {T.}~\bibnamefont {Shimojima}}, \bibinfo {author}
  {\bibfnamefont {T.}~\bibnamefont {Sonobe}}, \bibinfo {author} {\bibfnamefont
  {K.}~\bibnamefont {Koizumi}}, \bibinfo {author} {\bibfnamefont
  {S.}~\bibnamefont {Shin}}, \bibinfo {author} {\bibfnamefont {H.}~\bibnamefont
  {Miyahara}}, \bibinfo {author} {\bibfnamefont {A.}~\bibnamefont {Kimura}},
  \emph {et~al.},\ }\href
  {https://www.nature.com/nmat/journal/v10/n7/full/nmat3051.html} {\bibfield
  {journal} {\bibinfo  {journal} {Nat. Mater.}\ }\textbf {\bibinfo {volume}
  {10}},\ \bibinfo {pages} {521} (\bibinfo {year} {2011})}\BibitemShut
  {NoStop}%
\bibitem [{\citenamefont {Eremeev}\ \emph {et~al.}(2012)\citenamefont
  {Eremeev}, \citenamefont {Nechaev}, \citenamefont {Koroteev}, \citenamefont
  {Echenique},\ and\ \citenamefont {Chulkov}}]{eremeev2012ideal}%
  \BibitemOpen
  \bibfield  {author} {\bibinfo {author} {\bibfnamefont {S.~V.}\ \bibnamefont
  {Eremeev}}, \bibinfo {author} {\bibfnamefont {I.~A.}\ \bibnamefont
  {Nechaev}}, \bibinfo {author} {\bibfnamefont {Y.~M.}\ \bibnamefont
  {Koroteev}}, \bibinfo {author} {\bibfnamefont {P.~M.}\ \bibnamefont
  {Echenique}}, \ and\ \bibinfo {author} {\bibfnamefont {E.~V.}\ \bibnamefont
  {Chulkov}},\ }\href
  {https://journals.aps.org/prl/abstract/10.1103/PhysRevLett.108.246802}
  {\bibfield  {journal} {\bibinfo  {journal} {Phys. Rev. Lett.}\ }\textbf
  {\bibinfo {volume} {108}},\ \bibinfo {pages} {246802} (\bibinfo {year}
  {2012})}\BibitemShut {NoStop}%
\bibitem [{\citenamefont {Varykhalov}\ \emph {et~al.}(2012)\citenamefont
  {Varykhalov}, \citenamefont {Marchenko}, \citenamefont {Scholz},
  \citenamefont {Rienks}, \citenamefont {Kim}, \citenamefont {Bihlmayer},
  \citenamefont {S{\'a}nchez-Barriga},\ and\ \citenamefont
  {Rader}}]{varykhalov2012ir}%
  \BibitemOpen
  \bibfield  {author} {\bibinfo {author} {\bibfnamefont {A.}~\bibnamefont
  {Varykhalov}}, \bibinfo {author} {\bibfnamefont {D.}~\bibnamefont
  {Marchenko}}, \bibinfo {author} {\bibfnamefont {M.}~\bibnamefont {Scholz}},
  \bibinfo {author} {\bibfnamefont {E.}~\bibnamefont {Rienks}}, \bibinfo
  {author} {\bibfnamefont {T.}~\bibnamefont {Kim}}, \bibinfo {author}
  {\bibfnamefont {G.}~\bibnamefont {Bihlmayer}}, \bibinfo {author}
  {\bibfnamefont {J.}~\bibnamefont {S{\'a}nchez-Barriga}}, \ and\ \bibinfo
  {author} {\bibfnamefont {O.}~\bibnamefont {Rader}},\ }\href
  {https://journals.aps.org/prl/abstract/10.1103/PhysRevLett.108.066804}
  {\bibfield  {journal} {\bibinfo  {journal} {Phy. Rev. Lett.}\ }\textbf
  {\bibinfo {volume} {108}},\ \bibinfo {pages} {066804} (\bibinfo {year}
  {2012})}\BibitemShut {NoStop}%
\bibitem [{\citenamefont {Di~Sante}\ \emph {et~al.}(2013)\citenamefont
  {Di~Sante}, \citenamefont {Barone}, \citenamefont {Bertacco},\ and\
  \citenamefont {Picozzi}}]{di2013electric}%
  \BibitemOpen
  \bibfield  {author} {\bibinfo {author} {\bibfnamefont {D.}~\bibnamefont
  {Di~Sante}}, \bibinfo {author} {\bibfnamefont {P.}~\bibnamefont {Barone}},
  \bibinfo {author} {\bibfnamefont {R.}~\bibnamefont {Bertacco}}, \ and\
  \bibinfo {author} {\bibfnamefont {S.}~\bibnamefont {Picozzi}},\ }\href
  {http://onlinelibrary.wiley.com/doi/10.1002/adma.201203199/full} {\bibfield
  {journal} {\bibinfo  {journal} {Adv. Mater.}\ }\textbf {\bibinfo {volume}
  {25}},\ \bibinfo {pages} {509} (\bibinfo {year} {2013})}\BibitemShut
  {NoStop}%
\bibitem [{\citenamefont {Liebmann}\ \emph {et~al.}(2016)\citenamefont
  {Liebmann}, \citenamefont {Rinaldi}, \citenamefont {Di~Sante}, \citenamefont
  {Kellner}, \citenamefont {Pauly}, \citenamefont {Wang}, \citenamefont
  {Boschker}, \citenamefont {Giussani}, \citenamefont {Bertoli}, \citenamefont
  {Cantoni} \emph {et~al.}}]{liebmann2016giant}%
  \BibitemOpen
  \bibfield  {author} {\bibinfo {author} {\bibfnamefont {M.}~\bibnamefont
  {Liebmann}}, \bibinfo {author} {\bibfnamefont {C.}~\bibnamefont {Rinaldi}},
  \bibinfo {author} {\bibfnamefont {D.}~\bibnamefont {Di~Sante}}, \bibinfo
  {author} {\bibfnamefont {J.}~\bibnamefont {Kellner}}, \bibinfo {author}
  {\bibfnamefont {C.}~\bibnamefont {Pauly}}, \bibinfo {author} {\bibfnamefont
  {R.~N.}\ \bibnamefont {Wang}}, \bibinfo {author} {\bibfnamefont {J.~E.}\
  \bibnamefont {Boschker}}, \bibinfo {author} {\bibfnamefont {A.}~\bibnamefont
  {Giussani}}, \bibinfo {author} {\bibfnamefont {S.}~\bibnamefont {Bertoli}},
  \bibinfo {author} {\bibfnamefont {M.}~\bibnamefont {Cantoni}},  \emph
  {et~al.},\ }\href
  {http://onlinelibrary.wiley.com/doi/10.1002/adma.201503459/full} {\bibfield
  {journal} {\bibinfo  {journal} {Adv. Mater.}\ }\textbf {\bibinfo {volume}
  {28}},\ \bibinfo {pages} {560} (\bibinfo {year} {2016})}\BibitemShut
  {NoStop}%
\bibitem [{\citenamefont {Matetskiy}\ \emph {et~al.}(2015)\citenamefont
  {Matetskiy}, \citenamefont {Ichinokura}, \citenamefont {Bondarenko},
  \citenamefont {Tupchaya}, \citenamefont {Gruznev}, \citenamefont {Zotov},
  \citenamefont {Saranin}, \citenamefont {Hobara}, \citenamefont {Takayama},\
  and\ \citenamefont {Hasegawa}}]{matetskiy2015two}%
  \BibitemOpen
  \bibfield  {author} {\bibinfo {author} {\bibfnamefont {A.}~\bibnamefont
  {Matetskiy}}, \bibinfo {author} {\bibfnamefont {S.}~\bibnamefont
  {Ichinokura}}, \bibinfo {author} {\bibfnamefont {L.}~\bibnamefont
  {Bondarenko}}, \bibinfo {author} {\bibfnamefont {A.}~\bibnamefont
  {Tupchaya}}, \bibinfo {author} {\bibfnamefont {D.}~\bibnamefont {Gruznev}},
  \bibinfo {author} {\bibfnamefont {A.}~\bibnamefont {Zotov}}, \bibinfo
  {author} {\bibfnamefont {A.}~\bibnamefont {Saranin}}, \bibinfo {author}
  {\bibfnamefont {R.}~\bibnamefont {Hobara}}, \bibinfo {author} {\bibfnamefont
  {A.}~\bibnamefont {Takayama}}, \ and\ \bibinfo {author} {\bibfnamefont
  {S.}~\bibnamefont {Hasegawa}},\ }\href
  {https://journals.aps.org/prl/abstract/10.1103/PhysRevLett.115.147003}
  {\bibfield  {journal} {\bibinfo  {journal} {Phys. Rev. Lett.}\ }\textbf
  {\bibinfo {volume} {115}},\ \bibinfo {pages} {147003} (\bibinfo {year}
  {2015})}\BibitemShut {NoStop}%
\bibitem [{\citenamefont {Volobuev}\ \emph {et~al.}(2017)\citenamefont
  {Volobuev}, \citenamefont {Mandal}, \citenamefont {Galicka}, \citenamefont
  {Caha}, \citenamefont {S{\'a}nchez-Barriga}, \citenamefont {Di~Sante},
  \citenamefont {Varykhalov}, \citenamefont {Khiar}, \citenamefont {Picozzi},
  \citenamefont {Bauer} \emph {et~al.}}]{volobuev2017giant}%
  \BibitemOpen
  \bibfield  {author} {\bibinfo {author} {\bibfnamefont {V.~V.}\ \bibnamefont
  {Volobuev}}, \bibinfo {author} {\bibfnamefont {P.~S.}\ \bibnamefont
  {Mandal}}, \bibinfo {author} {\bibfnamefont {M.}~\bibnamefont {Galicka}},
  \bibinfo {author} {\bibfnamefont {O.}~\bibnamefont {Caha}}, \bibinfo {author}
  {\bibfnamefont {J.}~\bibnamefont {S{\'a}nchez-Barriga}}, \bibinfo {author}
  {\bibfnamefont {D.}~\bibnamefont {Di~Sante}}, \bibinfo {author}
  {\bibfnamefont {A.}~\bibnamefont {Varykhalov}}, \bibinfo {author}
  {\bibfnamefont {A.}~\bibnamefont {Khiar}}, \bibinfo {author} {\bibfnamefont
  {S.}~\bibnamefont {Picozzi}}, \bibinfo {author} {\bibfnamefont
  {G.}~\bibnamefont {Bauer}},  \emph {et~al.},\ }\href
  {http://onlinelibrary.wiley.com/doi/10.1002/adma.201604185/full} {\bibfield
  {journal} {\bibinfo  {journal} {Adv. Mater.}\ }\textbf {\bibinfo {volume}
  {29}},\ \bibinfo {pages} {1604185} (\bibinfo {year} {2017})}\BibitemShut
  {NoStop}%
\bibitem [{\citenamefont {Niesner}\ \emph {et~al.}(2016)\citenamefont
  {Niesner}, \citenamefont {Wilhelm}, \citenamefont {Levchuk}, \citenamefont
  {Osvet}, \citenamefont {Shrestha}, \citenamefont {Batentschuk}, \citenamefont
  {Brabec},\ and\ \citenamefont {Fauster}}]{niesner2016giant}%
  \BibitemOpen
  \bibfield  {author} {\bibinfo {author} {\bibfnamefont {D.}~\bibnamefont
  {Niesner}}, \bibinfo {author} {\bibfnamefont {M.}~\bibnamefont {Wilhelm}},
  \bibinfo {author} {\bibfnamefont {I.}~\bibnamefont {Levchuk}}, \bibinfo
  {author} {\bibfnamefont {A.}~\bibnamefont {Osvet}}, \bibinfo {author}
  {\bibfnamefont {S.}~\bibnamefont {Shrestha}}, \bibinfo {author}
  {\bibfnamefont {M.}~\bibnamefont {Batentschuk}}, \bibinfo {author}
  {\bibfnamefont {C.}~\bibnamefont {Brabec}}, \ and\ \bibinfo {author}
  {\bibfnamefont {T.}~\bibnamefont {Fauster}},\ }\href
  {https://journals.aps.org/prl/abstract/10.1103/PhysRevLett.117.126401}
  {\bibfield  {journal} {\bibinfo  {journal} {Phys. Rev. Lett.}\ }\textbf
  {\bibinfo {volume} {117}},\ \bibinfo {pages} {126401} (\bibinfo {year}
  {2016})}\BibitemShut {NoStop}%
\bibitem [{\citenamefont {Kane}\ and\ \citenamefont
  {Mele}(2005)}]{KanePRL2005}%
  \BibitemOpen
  \bibfield  {author} {\bibinfo {author} {\bibfnamefont {C.~L.}\ \bibnamefont
  {Kane}}\ and\ \bibinfo {author} {\bibfnamefont {E.~J.}\ \bibnamefont
  {Mele}},\ }\href {\doibase 10.1103/PhysRevLett.95.226801} {\bibfield
  {journal} {\bibinfo  {journal} {Phys. Rev. Lett.}\ }\textbf {\bibinfo
  {volume} {95}},\ \bibinfo {pages} {226801} (\bibinfo {year}
  {2005})}\BibitemShut {NoStop}%
\bibitem [{\citenamefont {Bercioux}\ and\ \citenamefont
  {De~Martino}(2010)}]{Bercioux2010}%
  \BibitemOpen
  \bibfield  {author} {\bibinfo {author} {\bibfnamefont {D.}~\bibnamefont
  {Bercioux}}\ and\ \bibinfo {author} {\bibfnamefont {A.}~\bibnamefont
  {De~Martino}},\ }\href {\doibase 10.1103/PhysRevB.81.165410} {\bibfield
  {journal} {\bibinfo  {journal} {Phys. Rev. B}\ }\textbf {\bibinfo {volume}
  {81}},\ \bibinfo {pages} {165410} (\bibinfo {year} {2010})}\BibitemShut
  {NoStop}%
\bibitem [{\citenamefont {Manchon}\ \emph {et~al.}(2015)\citenamefont
  {Manchon}, \citenamefont {Koo}, \citenamefont {Nitta}, \citenamefont
  {Frolov},\ and\ \citenamefont {Duine}}]{Manchon}%
  \BibitemOpen
  \bibfield  {author} {\bibinfo {author} {\bibfnamefont {A.}~\bibnamefont
  {Manchon}}, \bibinfo {author} {\bibfnamefont {H.~C.}\ \bibnamefont {Koo}},
  \bibinfo {author} {\bibfnamefont {J.}~\bibnamefont {Nitta}}, \bibinfo
  {author} {\bibfnamefont {S.~M.}\ \bibnamefont {Frolov}}, \ and\ \bibinfo
  {author} {\bibfnamefont {R.~A.}\ \bibnamefont {Duine}},\ }\href {\doibase
  10.1038/nmat4360} {\bibfield  {journal} {\bibinfo  {journal} {Nat Mater}\
  }\textbf {\bibinfo {volume} {14}},\ \bibinfo {pages} {871} (\bibinfo {year}
  {2015})}\BibitemShut {NoStop}%
\bibitem [{\citenamefont {Sheng}\ \emph {et~al.}(2017)\citenamefont {Sheng},
  \citenamefont {Yu}, \citenamefont {Yu}, \citenamefont {Weng},\ and\
  \citenamefont {Yang}}]{Sheng_JPCL}%
  \BibitemOpen
  \bibfield  {author} {\bibinfo {author} {\bibfnamefont {X.-L.}\ \bibnamefont
  {Sheng}}, \bibinfo {author} {\bibfnamefont {Z.-M.}\ \bibnamefont {Yu}},
  \bibinfo {author} {\bibfnamefont {R.}~\bibnamefont {Yu}}, \bibinfo {author}
  {\bibfnamefont {H.}~\bibnamefont {Weng}}, \ and\ \bibinfo {author}
  {\bibfnamefont {S.~A.}\ \bibnamefont {Yang}},\ }\href {\doibase
  10.1021/acs.jpclett.7b01390} {\bibfield  {journal} {\bibinfo  {journal}
  {Jour. Phys. Chem. Lett.}\ }\textbf {\bibinfo {volume} {8}},\ \bibinfo
  {pages} {3506} (\bibinfo {year} {2017})}\BibitemShut {NoStop}%
\bibitem [{\citenamefont {Souma}\ \emph {et~al.}(2011)\citenamefont {Souma},
  \citenamefont {Kosaka}, \citenamefont {Sato}, \citenamefont {Komatsu},
  \citenamefont {Takayama}, \citenamefont {Takahashi}, \citenamefont {Kriener},
  \citenamefont {Segawa},\ and\ \citenamefont {Ando}}]{Souma2011}%
  \BibitemOpen
  \bibfield  {author} {\bibinfo {author} {\bibfnamefont {S.}~\bibnamefont
  {Souma}}, \bibinfo {author} {\bibfnamefont {K.}~\bibnamefont {Kosaka}},
  \bibinfo {author} {\bibfnamefont {T.}~\bibnamefont {Sato}}, \bibinfo {author}
  {\bibfnamefont {M.}~\bibnamefont {Komatsu}}, \bibinfo {author} {\bibfnamefont
  {A.}~\bibnamefont {Takayama}}, \bibinfo {author} {\bibfnamefont
  {T.}~\bibnamefont {Takahashi}}, \bibinfo {author} {\bibfnamefont
  {M.}~\bibnamefont {Kriener}}, \bibinfo {author} {\bibfnamefont
  {K.}~\bibnamefont {Segawa}}, \ and\ \bibinfo {author} {\bibfnamefont
  {Y.}~\bibnamefont {Ando}},\ }\href {\doibase 10.1103/PhysRevLett.106.216803}
  {\bibfield  {journal} {\bibinfo  {journal} {Phys. Rev. Lett.}\ }\textbf
  {\bibinfo {volume} {106}},\ \bibinfo {pages} {216803} (\bibinfo {year}
  {2011})}\BibitemShut {NoStop}%
\bibitem [{\citenamefont {Sheng}\ and\ \citenamefont
  {Nikoli\ifmmode~\acute{c}\else \'{c}\fi{}}(2017)}]{Sheng2017}%
  \BibitemOpen
  \bibfield  {author} {\bibinfo {author} {\bibfnamefont {X.-L.}\ \bibnamefont
  {Sheng}}\ and\ \bibinfo {author} {\bibfnamefont {B.~K.}\ \bibnamefont
  {Nikoli\ifmmode~\acute{c}\else \'{c}\fi{}}},\ }\href {\doibase
  10.1103/PhysRevB.95.201402} {\bibfield  {journal} {\bibinfo  {journal} {Phys.
  Rev. B}\ }\textbf {\bibinfo {volume} {95}},\ \bibinfo {pages} {201402}
  (\bibinfo {year} {2017})}\BibitemShut {NoStop}%
\bibitem [{\citenamefont {Chen}\ \emph {et~al.}(2017)\citenamefont {Chen},
  \citenamefont {Wang}, \citenamefont {Liu}, \citenamefont {Yu}, \citenamefont
  {Sheng}, \citenamefont {Chen},\ and\ \citenamefont {Yang}}]{CongCaAgBi}%
  \BibitemOpen
  \bibfield  {author} {\bibinfo {author} {\bibfnamefont {C.}~\bibnamefont
  {Chen}}, \bibinfo {author} {\bibfnamefont {S.-S.}\ \bibnamefont {Wang}},
  \bibinfo {author} {\bibfnamefont {L.}~\bibnamefont {Liu}}, \bibinfo {author}
  {\bibfnamefont {Z.-M.}\ \bibnamefont {Yu}}, \bibinfo {author} {\bibfnamefont
  {X.-L.}\ \bibnamefont {Sheng}}, \bibinfo {author} {\bibfnamefont
  {Z.}~\bibnamefont {Chen}}, \ and\ \bibinfo {author} {\bibfnamefont {S.~A.}\
  \bibnamefont {Yang}},\ }\href
  {https://link.aps.org/doi/10.1103/PhysRevMaterials.1.044201} {\bibfield
  {journal} {\bibinfo  {journal} {Phys. Rev. Mat.}\ }\textbf {\bibinfo {volume}
  {1}},\ \bibinfo {pages} {044201} (\bibinfo {year} {2017})}\BibitemShut
  {NoStop}%
\bibitem [{\citenamefont {Zhang}\ \emph {et~al.}(2017)\citenamefont {Zhang},
  \citenamefont {Jin}, \citenamefont {Dai},\ and\ \citenamefont
  {Liu}}]{Zhang2017}%
  \BibitemOpen
  \bibfield  {author} {\bibinfo {author} {\bibfnamefont {X.}~\bibnamefont
  {Zhang}}, \bibinfo {author} {\bibfnamefont {L.}~\bibnamefont {Jin}}, \bibinfo
  {author} {\bibfnamefont {X.}~\bibnamefont {Dai}}, \ and\ \bibinfo {author}
  {\bibfnamefont {G.}~\bibnamefont {Liu}},\ }\href {\doibase
  10.1021/acs.jpclett.7b02129} {\bibfield  {journal} {\bibinfo  {journal} {J.
  Phys. Chem. Lett.}\ }\textbf {\bibinfo {volume} {8}},\ \bibinfo {pages}
  {4814} (\bibinfo {year} {2017})}\BibitemShut {NoStop}%
\bibitem [{\citenamefont {Li}\ and\ \citenamefont
  {Carbotte}(2014)}]{ZhouPRB2014}%
  \BibitemOpen
  \bibfield  {author} {\bibinfo {author} {\bibfnamefont {Z.}~\bibnamefont
  {Li}}\ and\ \bibinfo {author} {\bibfnamefont {J.~P.}\ \bibnamefont
  {Carbotte}},\ }\href {\doibase 10.1103/PhysRevB.89.165420} {\bibfield
  {journal} {\bibinfo  {journal} {Phys. Rev. B}\ }\textbf {\bibinfo {volume}
  {89}},\ \bibinfo {pages} {165420} (\bibinfo {year} {2014})}\BibitemShut
  {NoStop}%
\bibitem [{\citenamefont {Rycerz}\ \emph {et~al.}(2007)\citenamefont {Rycerz},
  \citenamefont {Tworzydlo},\ and\ \citenamefont
  {Beenakker}}]{rycerz2007valley}%
  \BibitemOpen
  \bibfield  {author} {\bibinfo {author} {\bibfnamefont {A.}~\bibnamefont
  {Rycerz}}, \bibinfo {author} {\bibfnamefont {J.}~\bibnamefont {Tworzydlo}}, \
  and\ \bibinfo {author} {\bibfnamefont {C.}~\bibnamefont {Beenakker}},\ }\href
  {http://www.nature.com/nphys/journal/v3/n3/full/nphys547.html?foxtrotcallback=true}
  {\bibfield  {journal} {\bibinfo  {journal} {Nat. Phys.}\ }\textbf {\bibinfo
  {volume} {3}},\ \bibinfo {pages} {172} (\bibinfo {year} {2007})}\BibitemShut
  {NoStop}%
\bibitem [{\citenamefont {Wu}\ \emph {et~al.}(2016)\citenamefont {Wu},
  \citenamefont {Liu}, \citenamefont {Chen}, \citenamefont {Xiao},\ and\
  \citenamefont {Liu}}]{wu2016full}%
  \BibitemOpen
  \bibfield  {author} {\bibinfo {author} {\bibfnamefont {Q.-P.}\ \bibnamefont
  {Wu}}, \bibinfo {author} {\bibfnamefont {Z.-F.}\ \bibnamefont {Liu}},
  \bibinfo {author} {\bibfnamefont {A.-X.}\ \bibnamefont {Chen}}, \bibinfo
  {author} {\bibfnamefont {X.-B.}\ \bibnamefont {Xiao}}, \ and\ \bibinfo
  {author} {\bibfnamefont {Z.-M.}\ \bibnamefont {Liu}},\ }\href
  {https://www.ncbi.nlm.nih.gov/pmc/articles/PMC4761925/} {\bibfield  {journal}
  {\bibinfo  {journal} {Sci. Rep.}\ }\textbf {\bibinfo {volume} {6}},\ \bibinfo
  {pages} {21590} (\bibinfo {year} {2016})}\BibitemShut {NoStop}%
\end{thebibliography}%

\end{document}